\documentclass{article} % For LaTeX2e
\usepackage{iclr2025_conference,times}

\usepackage{amsmath,amsfonts,bm}

\def\eqref#1{equation~\ref{#1}}
\def\1{\bm{1}}

\DeclareMathAlphabet{\mathsfit}{\encodingdefault}{\sfdefault}{m}{sl}
\SetMathAlphabet{\mathsfit}{bold}{\encodingdefault}{\sfdefault}{bx}{n}

\usepackage{hyperref}
\usepackage{url}
\usepackage{graphicx}
\usepackage{booktabs}
\usepackage{multirow}
\usepackage{algorithm}
\usepackage{algorithmic}
\usepackage{graphicx}
\usepackage{subcaption}

\title{Onboard Terrain Classification via Stacked Intelligent Metasurface-Diffractive Deep Neural Networks from SAR Level-0 Raw  Data }

\iclrfinalcopy

\author{Mengbing Liu \quad Xin Li \quad Jiancheng An \quad  Chau Yuen \\
{\small School of Electrical and Electronics Engineering (EEE)}\\
{\small Nanyang Technological University, Singapore} \\
{\tt\small \{mengbing001, xin019\}@e.ntu.edu.sg} \\
{\tt\small \{jiancheng.an, chau.yuen\}@ntu.edu.sg}
}
\begin{document}

\maketitle

\begin{abstract}
 
This paper introduces a novel approach for real-time onboard terrain classification from Sentinel-1 (S1) level-0 raw In-phase/Quadrature (IQ) data, leveraging a Stacked Intelligent Metasurface (SIM) to perform inference directly in the analog wave domain. Unlike conventional digital deep neural networks, the proposed multi-layer Diffractive Deep Neural Network (D$^2$NN) setup implements automatic feature extraction as electromagnetic waves propagate through stacked metasurface layers. This design not only reduces reliance on expensive downlink bandwidth and high-power computing at terrestrial stations but also achieves performance levels around \textbf{90\%} directly from the real raw IQ data, in terms of accuracy, precision, recall, and F1 Score. Our method therefore helps bridge the gap between next-generation remote sensing tasks and in-orbit processing needs, paving the way for computationally efficient remote sensing applications.

\end{abstract}

\section{Introduction}

Space-borne remote sensing missions increasingly rely on Synthetic Aperture Radar (SAR) data to support environmental and societal applications such as deforestation detection, flood monitoring, and agricultural assessment \citep{ haensch2010complex,zhang2017complex,ley2018exploiting,tottrup2022surface}. However, the continuous growth in data volume poses significant challenges in terms of downlink bandwidth, energy consumption, and real-time processing. Traditional pipelines typically transmit high-level products (e.g., SAR images) to terrestrial stations, incurring latency and cost \citep{filipponi2019sentinel}. To address these limitations, there is a pressing need for onboard classification solutions that operate \emph{at the source}, greatly reducing the downlink bandwidth requirements in sending data from satellite to terrestrial station, and enabling real-time, in-orbit decision-making.

Recent advances in deep learning have driven breakthroughs in land-cover classification, target detection, and image reconstruction from SAR data \citep{liu2024cromss,amieva2024super}. However, these methods often rely on digital backends, which remain constrained by onboard computational capabilities and energy budgets~\citep{boser1991analog}.  

Multi-layer diffractive metasurface arrays have recently emerged as a powerful analog architecture, implementing a layer-by-layer transformation on the propagating electromagnetic waves \citep{lin2018all, mengu2019analysis, liu2022programmable}. These so-called \emph{Stacked Intelligent Metasurfaces} (SIM) exhibit enhanced feature extraction capabilities by providing multiple programmable phase and amplitude modifications. Prior research has demonstrated these multi-layer metasurface networks can function as ultra-fast, energy-efficient inference engines \citep{an2023stacked, huang2024stacked,liu2024stacked}, but their application to \emph{onboard terrain classification} of   Sentinel-1 (S1) level-0 raw data has remained largely unexplored.

In this work, we propose a wave-based diffractive deep neural network (D$^2$NN) framework that processes S1 level-0 raw IQ data in orbit, reducing the need for extensive data transmission and computing resources at terrestrial stations. Our main contributions are:

\begin{itemize}
    \item \textbf{Multi-Layer Metasurface Inference:} We design a multi-layer SIM-based approach that performs layer-by-layer feature extraction in the electromagnetic domain, achieving higher representational capacity than single-layer systems. By leveraging analog wave propagation and straightforward data augmentation techniques, our method not only enhances the efficiency but also reduces the computational load and bandwidth requirements.
    \item \textbf{Enhancement through Data Augmentation:}
  Our research indicates that omitting phase-rotation data augmentation results in a substantial decrease in the F1 Score (from 90.60\% to 69.35\%). This highlights the essential role of data augmentation strategies in mitigating the inherent noise and Doppler effects in raw  level-0 data.
    \item \textbf{Real-World Terrain Classification:} We validate our model on actual S1 level-0 raw IQ data, achieving comprehensive performance levels around \textbf{90\%}, covering accuracy, precision, recall, and F1 Score, which facilitates  \emph{near real-time} terrain classification tasks.

\end{itemize}

\begin{figure}[t]
\begin{center}
    \includegraphics[width=0.8\textwidth]{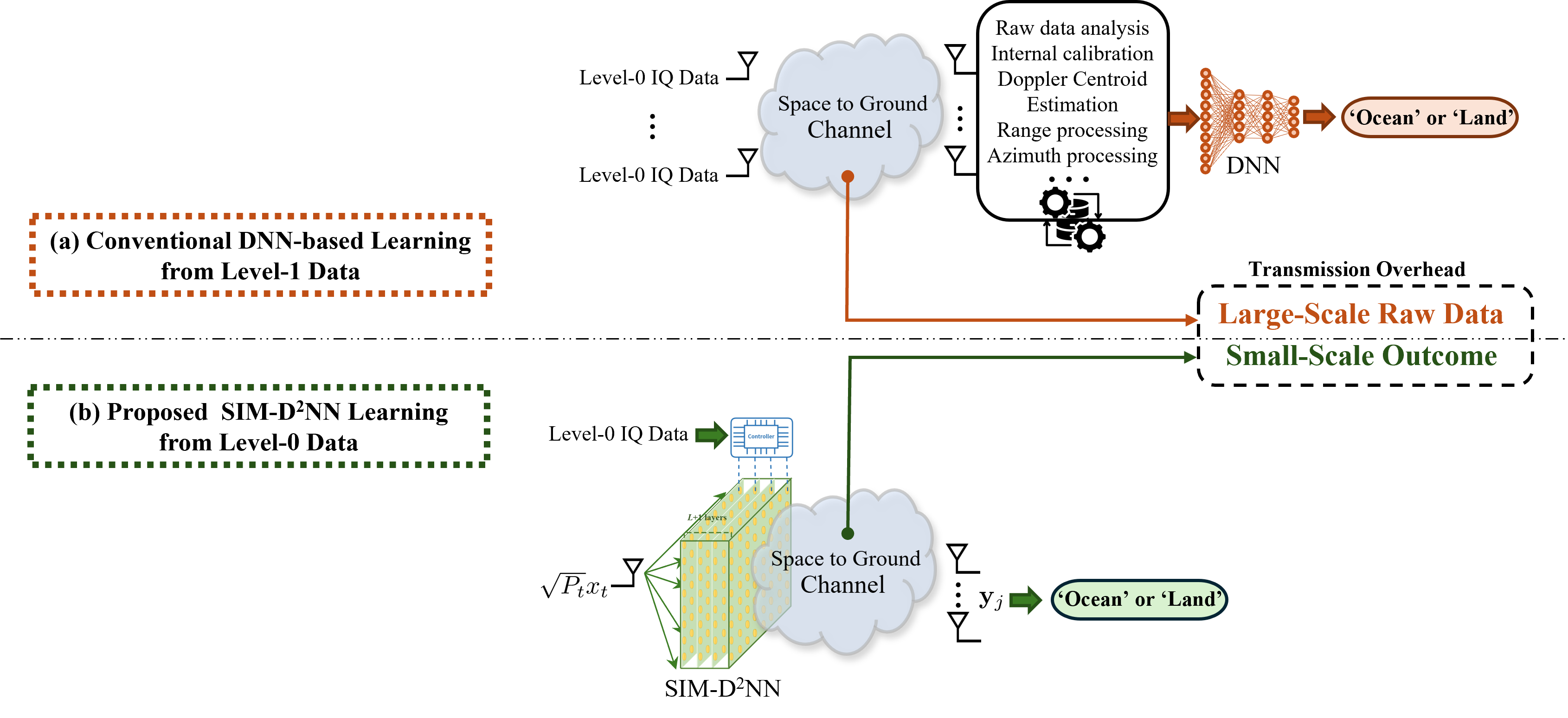}
\end{center}
\caption{\textbf{High-level system overview.}
    Comparison of processing progress: Traditional DNN-based learning from level-1 data vs. proposed D$^2$NN-based learning from level-0 data.}
    \label{system}
\end{figure}

\section{Methodology}
\label{headings}
\subsection{System Overview}

Figure~\ref{system} compares the conventional and proposed pipelines for terrain classification using SAR data. The conventional pipeline in Figure~\ref{system} (a) involves transmitting a large volume of raw data for downstream DNN processing. This is due to the complex algorithms required to convert level-0 IQ data into level-1 data, such as Single Look Complex (SLC) images, which cannot be processed onboard. Consequently, a large amount of the raw data needs to be transmitted and received between the satellite and the terrestrial station.

Conversely, the proposed method illustrated in Figure~\ref{system}~(b) processes level-0 IQ data directly. The input data is mapped onto the transmission coefficient pattern of the initial layer (0-th layer). 
Let $\sqrt{P_t}x_t$ represeal, with $P_t$ being the transmit power and $x_t$ the normalized electromagnetic signal such that $|x_t|^2 = 1$.
As the signal $\sqrt{P_t}x_t$ propagates through the initial layer, it carries the encoded input data to the subsequent layers (D$^2$NN enabled by SIM), where feature extraction and onboard classification are automatically performed. This setup requires only receiving merely the classification outcomes with a small amount of data, thereby streamlining the overall data processing workflow.

\subsection{The Architecture of SIM-D$^2$NN}
 
\paragraph{Layer-by-Layer Diffractive Network.}
We construct an $(L+1)$-layer SIM, as shown in Figure~\ref{system} (b), with each layer consisting of $M$ programmable meta-atoms. The 0-th layer $\mathbf{\Phi}_j^0 $ elements are reconfigured using a controller to align with the   augmented input features of the $j$-th patch $\boldsymbol{\overline{s}}_{j}$, enabling efficient manipulation of electromagnetic waves in the wave domain:
\begin{equation}
  \mathbf{\Phi}_j^0 = \mathrm{diag} \left( a_{j,1}^{0}e^{j\theta_{j,1}^{0}}, a_{j,2}^{0}e^{j\theta_{j,2}^{0}}, \ldots, a_{j,M}^{0}e^{j\theta_{j,M}^{0}} \right) = \mathrm{diag} \left( \overline{{s}}_{j,1}, \overline{{s}}_{j,2}, \ldots, \overline{{s}}_{j,M} \right).
\end{equation}
where $\overline{{s}}_{j,m}$ is the $m$-th element in the normalized input tensor after data augmentation. The subsequent layers $\mathbf{\Phi}^{l},   l \in \{1,2,\ldots, L\}$ apply learned phase shifts $\theta_{m}^l$ to the incoming wave to enable deep feature extraction. The diffracted wave from the final layer propagates to a $K$-element antenna array at the terrestrial station, where $K$ corresponds to the number of terrain classification categories.

\paragraph{Wave Propagation Model.}

Let \(\mathbf{W}^{l} \in \mathbb{C}^{M \times M}\) denote the transmission matrix between the \((l-1)\)-th and \(l\)-th metasurface layers, and \(\mathbf{w}^{0} \in \mathbb{C}^{M \times 1}\) be the vector from the transmit antenna to the $0$-th layer. Based on \citet{lin2018all}, the \((m,m')\)-th element \(w_{m,m'}^{l}\) is
\begin{equation}
w_{m,m'}^{l}
= \frac{d_x d_y \cos \chi_{m,m'}}{d_{m,m'}^{l}}
\left(\frac{1}{2\pi d_{m,m'}^{l}} - j\frac{1}{\lambda}\right)
e^{j \frac{2\pi d_{m,m'}^{l}}{\lambda}},
\label{eq:W_l}
\end{equation}
where \(\lambda\) is the wavelength, \(d_{m,m'}^{l}\) the distance between two meta-atoms considered, \(\chi_{m,m'}\) the propagation angle, and \(d_x \times d_y\) the meta-atom dimensions. Thus, the overall propagation matrix \(\mathbf{G}_j \in \mathbb{C}^{M \times M}\) is then 
$\mathbf{G}_j
= \mathbf{\Phi}^L
\mathbf{W}^L
\mathbf{\Phi}^{L-1}
\cdots
\mathbf{\Phi}^{1}
\mathbf{W}^{1}
\mathbf{\Phi}^{0}_j.$

\paragraph{Inference from the Received Signal.} The $\mathbf{y}_j \in \mathbb{C}^{K \times 1}$ received at the terrestrial station is
\begin{equation}
    \mathbf{y}_j = \mathbf{H}\,\mathbf{G}_j\,\mathbf{w}^0\,\sqrt{P_t}x_t + \mathbf{n},
\end{equation}
where $\mathbf{n} \sim \mathcal{CN}(0, \sigma^2)$ is the i.i.d. Additive White Gaussian Noise (AWGN), and $\mathbf{H} \in \mathbb{C}^{K \times M}$ models the channel from the SIM to the terrestrial station. After receiving $\mathbf{y}_j$ from $K$ antennas, the magnitude at antenna $k$ indicates the likelihood that the $j$-th patch belongs to category $k$. Classification is done by selecting the antenna with the highest signal magnitude, i.e., 
$\hat{k}_j
= \arg\max_{k \in \mathcal{K}}
\left\{
|y_{j,1}|^2,\,
|y_{j,2}|^2,\dots,
|y_{j,K}|^2
\right\},$
where $\hat{k}_j$ is the predicted category for patch $j$, and $\mathcal{K}$ is the set of all categories.

\section{Experiments and Discussion}
\label{others}
\subsection{Data Generation and Preparation}

To demonstrate the effectiveness of our proposed SIM-D\(^2\)NN, we classify land or ocean on S1 level-0 raw IQ data. By targeting this basic level of remote sensing data, we aim to highlight the practical feasibility of SIM-D\(^2\)NN for airborne classification in near real-time in actual satellite operations.

\textbf{Data Description.} We leverage S1 level-0 raw IQ data, partitioning the scene into $128 \times 128$ patches with a stride of 32 \citep{filipponi2019sentinel}. Unlike higher-level SAR products, level-0 IQ data preserve the original phase and amplitude information but lack standard radiometric and geometric corrections.

\textbf{Labeling Ground Truth.} Level-0 data lack direct annotations, making ground truth generation difficult. We employ an open-source S1 level-0 decoding algorithm to partially denoise and clarify the data \citep{hall2023sentinel1level0decoding}. We then annotate land vs. ocean regions using the decoded imagery as a reference for the raw IQ patches.

\textbf{Data Augmentation.} To mitigate the pervasive speckle noise in SAR imagery, we apply phase rotation to the raw data. Each patch undergoes a controlled phase shift by a predetermined angle and then is concatenated with the original raw data  \footnote{The input metasurface layer is divided into two halves, with one half configured based on the input data and the other half based on a 90-degree rotated version of the same data. The modulation occurs naturally as the carrier waves pass through the input layer.
}.
 This process enhances both the robustness and overall quality of inputs used for model training.

\subsection{Quantitative Evaluation}

We evaluate the performance of the proposed SIM-D$^2$NN compared to a digital DNN with the same architecture. Unlike SIM-D$^2$NN where weights are constrained to unit modulus for phase-only adjustments, the DNN allows unconstrained weight values, potentially leading to better performance.
In addition, ablation studies are conducted to assess critical design factors. Table~\ref{tab:comparison_methods} reports Precision, Recall, F1 score, and Overall Accuracy on the real S1 level-0 dataset. The training details, SIM architecture, and visualization results are in the Appendix.

\subsubsection{Main Comparisons}

 Table~\ref{tab:comparison_methods} compares the classification result of the SIM-D\(^2\)NN to a fully digital DNN.
Despite the inherent constraints associated with phase-only modulation in metasurface-based systems, the SIM-D\(^2\)NN attains a performance level of approximately 90\%, as measured by accuracy, precision, recall, and F1 Score. Remarkably, this performance is within a narrow margin of 5–7\%  of the digital DNN.

\subsubsection{Ablation Studies}

Table~\ref{tab:comparison_methods} shows ablation experiments to highlight important system parameters:

\textbf{Number of Metasurface Layers for Feature Extraction ($L$).} Increasing \( L \) enhances performance, with a precision score reaching 90.54\% at \( L=4 \), which surpasses the 87.63\% achieved at \( L=1 \).

\textbf{Transmit Power ($P_t$).} Reducing $P_t$ from $20$~dBm to $5$~dBm degrades accuracy to around $80\%$, reflecting the wave-domain model’s dependence on a sufficient Signal-to-Noise Ratio (SNR) to overcome speckle noise and channel variations.

\textbf{Sampling Rate ($S$).} Using $10\%$ of the total patches for training achieves near-optimal performance, showing that the SIM-D\(^2\)NN can learn efficient representations with limited training data. 

\textbf{Phase Rotation and Data Augmentation.} Omitting the phase-rotation augmentation leads to a significant drop in F1 (69.35\% vs. 90.60\%). This result underscores the importance of data augmentation in mitigating the random phase fluctuations inherent to SAR imagery.

\begin{table}[!ht]
\caption{Comparison of different scenarios on the S1 level-0 raw IQ dataset.}
\begin{center}
\resizebox{0.9\textwidth}{!}{% Resize table to fit within the text width
\begin{tabular}{l||cccc}
\toprule
\multirow{2}{*}[-4pt]{\textbf{Ablation Setting}}& \multicolumn{4}{c}{\textbf{S1 Level-0 Raw IQ Dataset}} \\
\cmidrule{2-5}
 & \textbf{Precision (\%)} $\uparrow$ & \textbf{Recall (\%)} $\uparrow$ & \textbf{F1 Score (\%)} $\uparrow$ & \textbf{Overall Accuracy (\%)} $\uparrow$  \\
\midrule
SIM-D$^2$NN ($L = 1$) & 87.63 & 91.27 & 89.41 & 83.44 \\
SIM-D$^2$NN ($L = 6$)  & 87.21 & 92.87 & 89.95& 88.15\\
SIM-D$^2$NN ($S = 5\%$) & 87.84 & 91.49 & 89.62 & 85.75 \\
SIM-D$^2$NN ($S = 20\%$) & 91.56 &93.98&92.76 & 89.31  \\
SIM-D$^2$NN ($P_t = 5 $ dBm) & 86.14 & 92.20 & 89.07 & 80.29 \\
SIM-D$^2$NN (No phase rotation) &  62.09 & 78.54 & 69.35 & 54.97 \\
SIM-D$^2$NN (Baseline)  & 90.54 & 90.67 &90.60 & 87.83 \\
\midrule
\textit{Digital DNN} & \textit{94.78} & \textit{97.14} & \textit{95.95} & \textit{92.91} \\
\bottomrule
\multicolumn{5}{l}{ \textbf{Note:} Our baseline SIM-D\(^2\)NN uses $L=4$ layers, $P_t=20$~dBm, and $S=10\%$.}
\end{tabular}
}
\end{center}
\label{tab:comparison_methods}
\end{table}

\subsection{Discussion}

The classification via SIM-D$^2$NN can significantly reduce data transmission overhead, enhancing applications such as near real-time flood detection and deforestation alerts. Future work will expand from the land/ocean classifications in this paper to a more comprehensive terrain classification task.

The approach relies on specialized metasurface hardware, which is constrained to linear operations and limits the SIM-D$^2$NN from performing critical nonlinear functions essential for boosting DNN performance. Additionally, while our simulations consider noise and phase modulus constraints, real-world communication links might introduce more complex distortions like time-varying channels and imperfect channel estimation.

\section{Conclusion}
We have developed a multi-layer, SIM-D$^2$NN designed to process S1 raw IQ data for terrain classification. By harnessing the inherent properties of wave propagation through multiple metasurface layers, this approach has demonstrated the ability to achieve as high a performance as around 90\%. This significant performance boost reduces dependence on digital processing backends and lowers the costs associated with data transmission. These encouraging outcomes underscore the potential of wave-domain analog computing to revolutionize remote sensing technologies, offering faster, more efficient, and sustainable solutions for Earth observation.

\section{Acknowledgment}
This research received funding from the Agency for Science, Technology and Research (A*STAR), Singapore, under Grant No. M22L1b0110. It was also supported by the National Research Foundation, Singapore, and the Infocomm Media Development Authority under the Future Communications Research \& Development Programme (FCP-NTU-RG-2024-025).

\bibliography{iclr2025_conference}
\bibliographystyle{iclr2025_conference}

\clearpage
\appendix
\section{Appendix}
\label{appendix:details}
\subsection{The architecture of SIM}
 \begin{figure}[th]
 \begin{center}
    \includegraphics[width=0.8\linewidth]{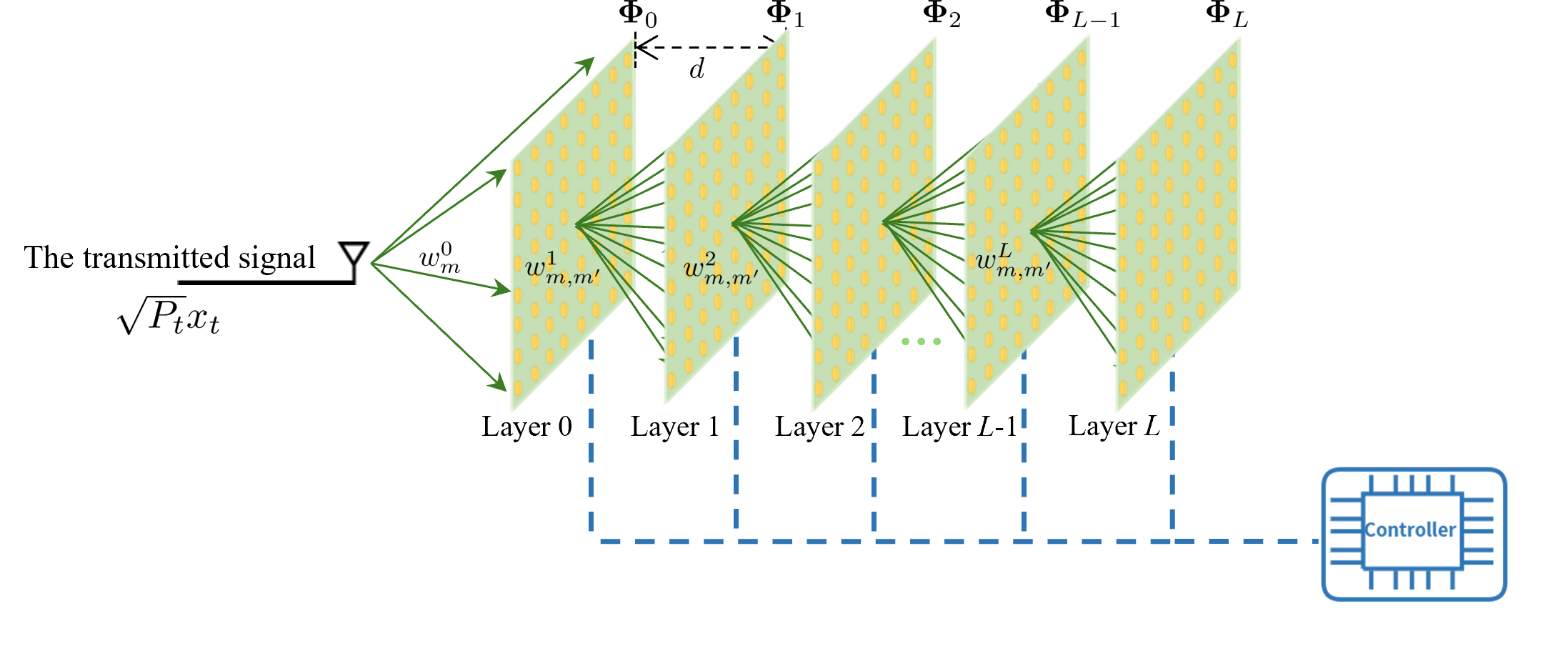}
 \end{center}
    \caption{The illustration of the SIM structure and corresponding parameters.}
    \label{SIM_architecture}
\end{figure}
 Figure~\ref{SIM_architecture} illustrates the high-level architecture and key parameters of the SIM. Each layer contains a grid of meta-atoms $\mathbf{\Phi}_l$ (for $l \in \{1,\ldots,L\}$), whose phase shifts $\theta$ are learnable. By stacking multiple layers in series, the SIM performs sequential waveform transformations. The interlayer spacing $d$ and the channel coefficient $w$ are fixed system parameters that shape diffraction and phase accumulation, while the learnable phase shifts $\theta$ effectively carry out the linear operations approximating certain DNN computations in the wave domain. Specifically, the normalized feature is applied at layer 0, so that when the signal $\sqrt{P_t}x_t$ passes through this layer, the feature is transmitted to subsequent layers for further extraction. 

\subsection{Training and Deployment Algorithm Framework}

\begin{algorithm}[htb]
\caption{Training and Deployment of the SIM-D$^2$NN}
\label{alg:Framwork}
\begin{algorithmic}[1]
\REQUIRE ~~\\
Sliding window patches $\mathcal{J} = \{{\boldsymbol{S}_{j}}\}_{j = 1}^{J}$;\\
Channel matrix $\mathbf{H}$,  $\mathbf{W}^{l}$ for $l \neq 0$, and $\mathbf{w}^{0}$.
\ENSURE ~~\\
S1 level-0 terrain classification result.

\STATE  \textbf{Stage 1: Offline Training}
\STATE  Randomly sample 10$\%$ of patches from $\mathcal{J}$. 
\FOR {epoch $= 1$ to $N_{e}$}
\STATE \( \overline{\mathbf{s}}_{j} \gets \text{Modulation}( {\mathbf{S}}_{j}) \) ;
\STATE Initialize \( \mathbf{\Phi}_j^0 \gets \text{diag}( \overline{s}_{j, 1},  \overline{s}_{j, 2}, \dots,  \overline{s}_{j, M}) \);
 
\STATE Compute $\mathbf{y}_{j}$ according to (3) and update the learnable parameters ${\mathbf{\theta}}_m^{l}$.
\ENDFOR
\RETURN Optimal phase shifts $\hat{\mathbf{\theta}}_m^{l}, \forall m \in \mathcal{M}, l \neq 0, l \in \mathcal{L}$.

\STATE  \textbf{Stage 2: SIM-Based D$^2$NN Deployment}
\FOR {each patch $j = 1$ to $J$}
\STATE Repeat steps 4-6 for feature embedding; 
\STATE Set the phase shift at each meta-atom using $\hat{\mathbf{\theta}}_m^{l}$; 
\STATE Compute $\mathbf{y}_j$  according to (3);
\STATE Classify by $\hat{k}_j = \arg\max\limits_{k \in \mathcal{C}} \left\{ |y_{j,1}|^2, |y_{j,2}|^2, \dots, |y_{j,C}|^2 \right\}$;
\ENDFOR
\RETURN  S1 level-0 terrain classification result.
\end{algorithmic}
\end{algorithm}
Algorithm~\ref{alg:Framwork} outlines the two-stage framework for training and deploying the SIM-aided communication system:
\begin{itemize}
    \item \textbf{Stage 1 (Offline Training):} A portion of the dataset (e.g., 10\%) is sampled to learn the optimal phase configurations that minimize classification loss. Each patch in the training set is downsampled and normalized, and its corresponding phase configuration matrix is initialized. The output signal is then computed, and the learnable phase parameters are updated via backpropagation.
    \item  \textbf{Stage 2 (SIM-D$^2$NN Deployment):} For each incoming patch, the same steps of modulation are performed, and the final learned phase shifts are applied to the meta-atoms at SIM. The received signal is probed and the antenna with the highest-intensity is selected to output the class.
\end{itemize}

\subsection{Simulation Parameters}
The system operates at a carrier frequency of 12 GHz, corresponding to a wavelength of \(\lambda = 25\) mm. The thickness of the SIM, \(T_{\rm SIM}\), is set to $0.05$ m, with the spacing between adjacent metasurfaces in an \(L\)-layer SIM defined as \(d_{L} =  {T_{\rm SIM}}/{L}\). Each meta-atom has dimensions of \(d_x = d_y =  {\lambda}/{2}\).
For the wireless link, unlike terrestrial communication scenarios, where channels in urban areas are typically modeled as Rayleigh fading, the space-to-ground channel is modeled using a Rician fading model, which accounts for both Line-of-Sight (LoS) and Non-Line-of-Sight (NLoS) components  with Rician factor of \(K = 20\) dB \citep{paulraj2003introduction}, and model the path loss from the SIM to the receiver as \citep{al2020modeling}:
\begin{equation}
    {\rm{PL}}(d, f) = {\rm{FSPL}}(d, f) + {\rm{LA}}  + {\rm{LE}},
\end{equation}
where \(f\) and \( d\) is the carrier frequency and the distance, respectively. \( {\rm{LA}} \) represents attenuation due to atmospheric absorption and \({\rm{LE}} \) characterizes path loss due to interactions with near-surface urban structures. \({\rm{FSPL}}(d, f)\) is the free space path loss expressed as
\begin{equation}
    {\rm{FSPL}}(d, f)= 20\log(f) + 20\log(d) - 147.55.
\end{equation}

Additionally, Table~\ref{tab:hyperparams} summarizes other key hyperparameters, such as metasurface size, number of layers, optimizer, learning rate, batch size, and noise levels. All experiments adhere to these parameters without any specialized configuration.

\begin{table}[h]
    \caption{Simulation and training hyperparameters}
\begin{center}
    \begin{tabular}{l l}
    \toprule
    \textbf{Parameter} & \textbf{Value} \\
    \midrule
    Metasurface size ($M$) & $2048$ \\
    Number of layers ($L$) & 4 \\
    Receive antenna elements ($K$)  & 2\\
    Transmit antenna elements & 1 \\
    Optimizer & AdamW \\
    Initial learning rate & 0.01 \\
    Batch size &  64\\
    Epochs & 60 \\
    Noise power & -104 dBm \\
    \bottomrule
    \end{tabular}
\end{center}
    \label{tab:hyperparams}
\end{table}

\begin{figure}
    \centering
    \begin{subfigure}[b]{0.4\textwidth}
        \centering
        \includegraphics[width=0.8\textwidth]{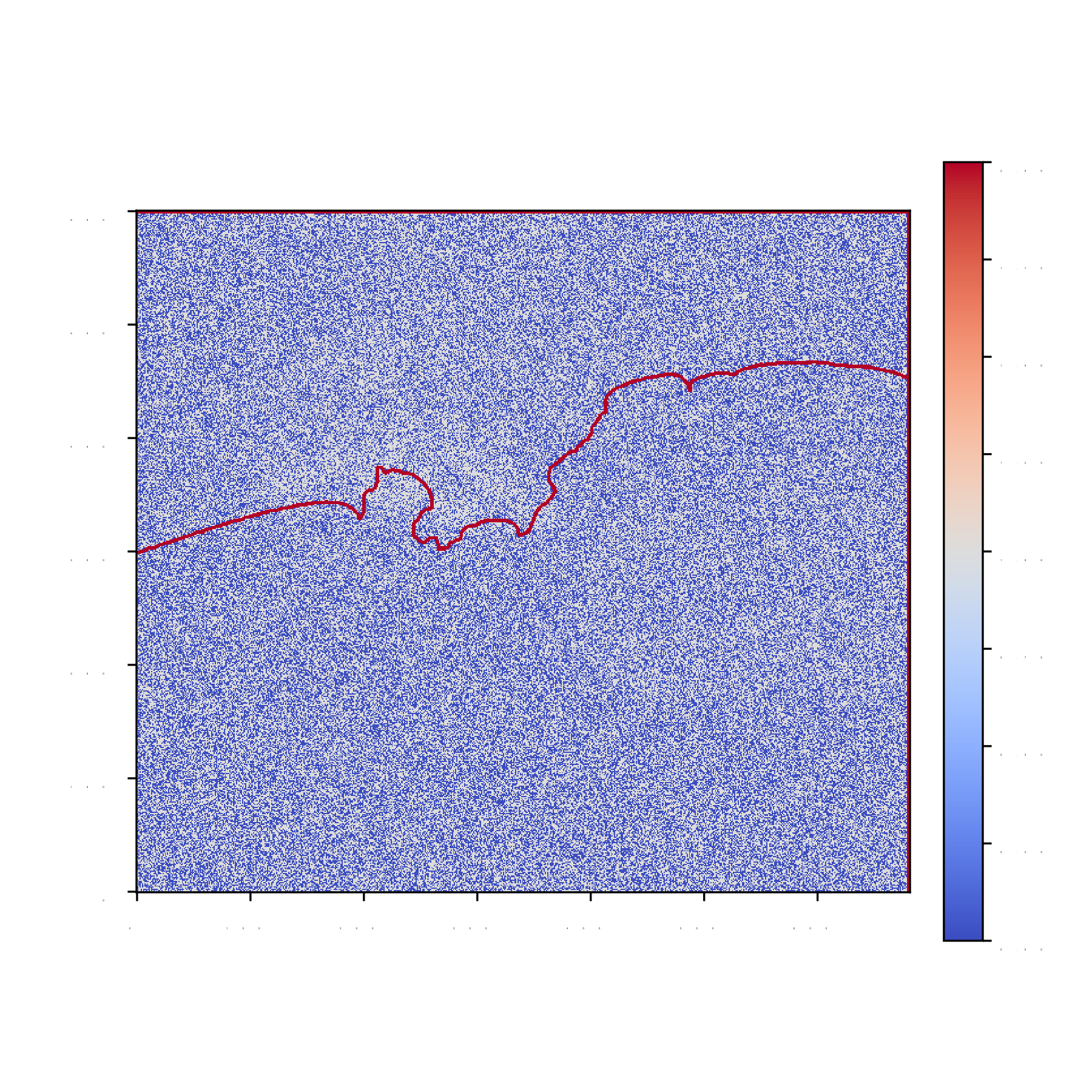}
        \caption{SIM-D$^2$NN (No phase rotation)}
        \label{fig:sub1}
    \end{subfigure}
    \begin{subfigure}[b]{0.4\textwidth}
        \centering
        \includegraphics[width=0.8\textwidth]{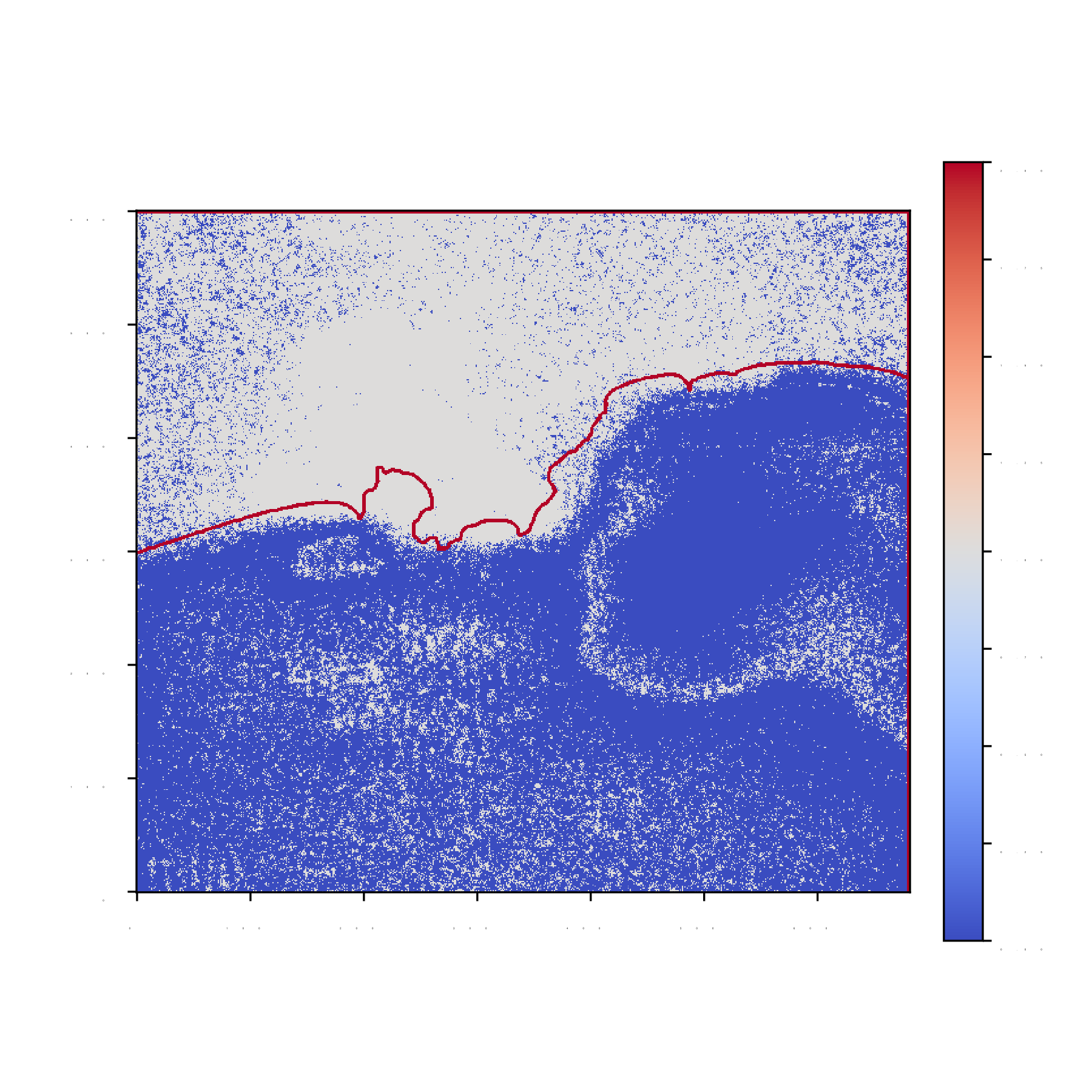}
        \caption{SIM-D$^2$NN (Baseline)}
        \label{fig:sub2}
    \end{subfigure}
    \\
    \begin{subfigure}[b]{0.4\textwidth}
        \centering
        \includegraphics[width=0.8\textwidth]{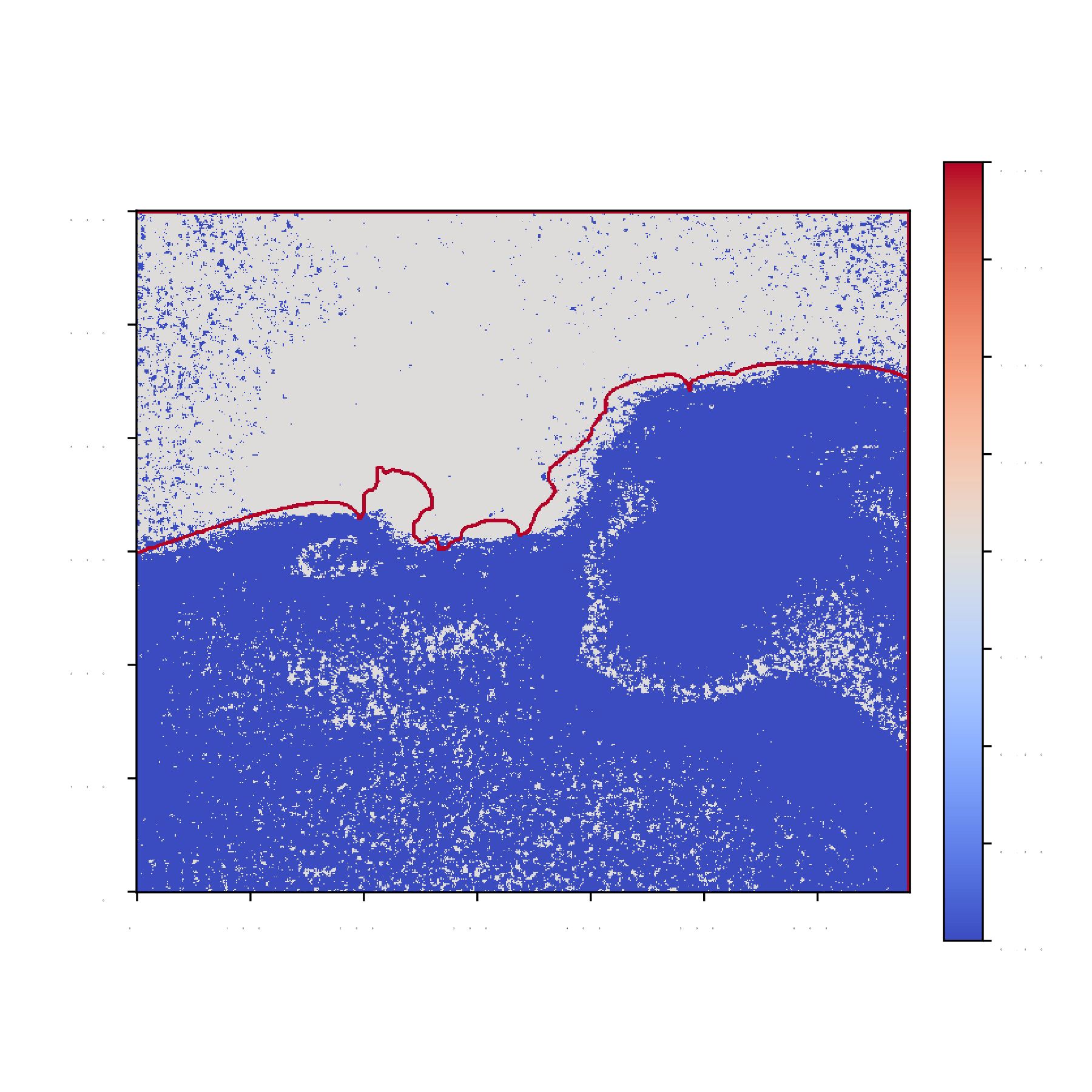}
        \caption{Digital DNN}
        \label{fig:sub3}
    \end{subfigure}
    \begin{subfigure}[b]{0.4\textwidth}
        \centering
        \includegraphics[width=0.8\textwidth]{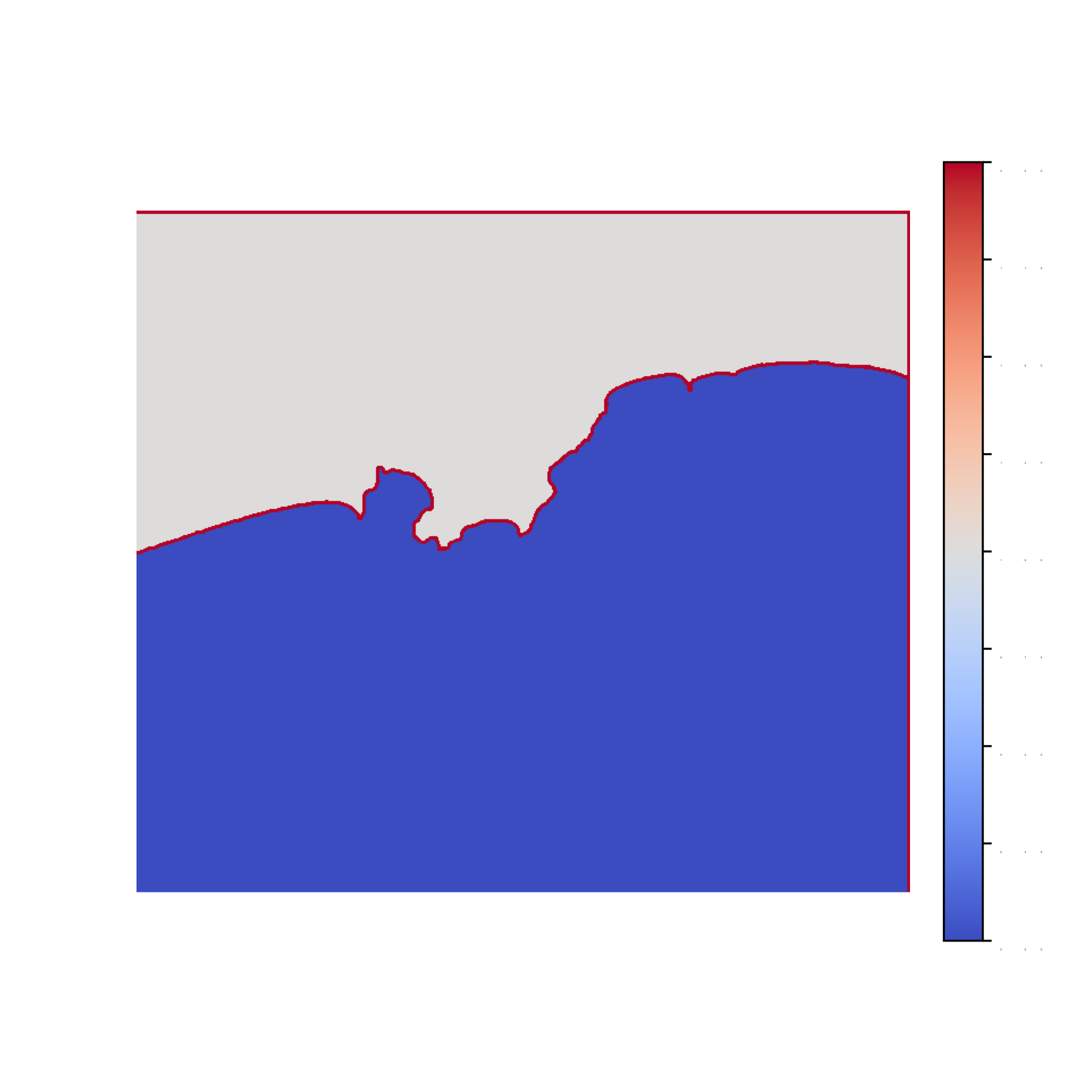}
        \caption{Ground Truth Label}
        \label{fig:sub4}
    \end{subfigure}
    \caption{Comparison of the visualization results under different methods.}
    \label{fig:overall}
\end{figure}

 \subsection{The visualization results on the whole dataset}

Figure \ref{fig:overall} displays the classification outcomes for the entire S1 level-0 dataset. Each patch is categorized based on the predicted terrain class, offering a detailed panorama of how different types of land covers—such as ocean or land—are represented. These visualizations facilitate the identification of areas susceptible to misclassification, providing critical insights that can be used to optimize the network architecture, hyperparameters, and data augmentation techniques. 

The comparison with the ground truth label, as shown in Figure \ref{fig:overall} (d), illustrates the effectiveness of various methods, including phase rotation (data augmentation), SIM-D$^2$NN, and digital DNN, in identifying terrain features. The inclusion of phase rotation in data augmentation proves essential for effectively learning from the IQ raw data, as demonstrated in Figure \ref{fig:overall} (a). Through an analysis of Figure \ref{fig:overall} (b) and Figure \ref{fig:overall} (c), it becomes clear that the analog SIM-D$^2$NN achieves results in terrain classification comparable to those obtained using a digital DNN. Notably, SIM-D$^2$NN manipulates only the phases at each meta-atom, adhering to unit modulus constraints, whereas the digital DNN operates without such limitations.

\end{document}